\documentclass[twocolumn,aps,pre,superscriptaddress]{revtex4}

\usepackage{amsmath}
\usepackage{graphicx}
\usepackage{MnSymbol}

\usepackage{color}
\usepackage[normalem]{ulem}

\begin{document}

\title{Finite-size scaling in the interfacial stiffness of
rough elastic contacts}

\author{Lars Pastewka}
\affiliation{Dept. of Physics and Astronomy, Johns Hopkins University, Baltimore, MD 21218, USA}
\affiliation{MikroTribologie Centrum $\mu$TC, Fraunhofer-Institut f\"ur Werkstoffmechanik IWM, 79108 Freiburg, Germany}
\author{Nikolay Prodanov}
\affiliation{J\"ulich Supercomputing Center, Institute for Advanced Simulation, FZ J\"ulich, 52425 J\"ulich, Germany}
\affiliation{Dept. of Materials Science and Engineering, Universit\"at des Saarlandes, 66123 Saarbr\"ucken, Germany}
\author{Boris Lorenz}
\affiliation{Peter Gr\"unberg Institut-1, FZ-J\"ulich, 52425 J\"ulich, Germany}
\author{Martin H. M\"user}
\affiliation{J\"ulich Supercomputing Center, Institute for Advanced Simulation, FZ J\"ulich, 52425 J\"ulich, Germany}
\affiliation{Dept. of Materials Science and Engineering, Universit\"at des Saarlandes, 66123 Saarbr\"ucken, Germany}
\author{Mark O. Robbins}
\affiliation{Dept. of Physics and Astronomy, Johns Hopkins University, Baltimore, MD 21218, USA}
\author{Bo N. J. Persson}
\affiliation{Peter Gr\"unberg Institut-1, FZ-J\"ulich, 52425 J\"ulich, Germany}

\begin{abstract}
The total elastic stiffness of two contacting bodies with a microscopically rough interface has an interfacial contribution $K$ that is entirely attributable to surface roughness.
A quantitative understanding of $K$ is important because it can dominate the total mechanical response and because it is proportional to the interfacial contributions to electrical and thermal conductivity in continuum theory.
Numerical simulations of the dependence of $K$ on the applied squeezing
pressure $p$ are presented for nominally flat elastic solids with a range
of surface roughnesses.
Over a wide range of $p$, $K$ rises linearly with $p$.
Sublinear power-law scaling is observed at small $p$, but
the simulations reveal
that this is a finite-size effect.
We derive accurate, analytical expressions for
the exponents and prefactors of this low-pressure scaling
of $K$ by extending the contact mechanics theory of Persson to
systems of finite size.
In agreement with our simulations, these expressions show that the onset of the low-pressure scaling regime moves to lower pressure as the system size increases.
\end{abstract}

\maketitle



\section{Introduction}

Two solids in mechanical contact tend to touch at only a
miniscule fraction of their apparent contact area $A_0$, because
their surfaces are microscopically
rough~\cite{bowden56,dieterich94,persson01,hyun04}.
This imperfect contact has profound implications for transmission
of charge, heat and forces through the interface.
The effect of the interface can be expressed in terms of
an interfacial stiffness or conductance that adds in series
with the bulk response of two solids with ideal flat surfaces
\cite{Berthoud98,Barber03}.
Improved theories of these interfacial contributions are important
because they frequently dominate
the total response of the system and are a strong function
of the normal force $F$ (or load) pushing the solids together.
In this paper we consider the scaling of stiffness with $F$
for nonadhesive self-affine rough surfaces.
The results are more generally applicable since
the shear and normal stiffness and electrical and heat
conductance are all proportional to each other
within linear-response continuum mechanics
~\cite{Barber03}.

In a pioneering experimental work, Berthoud and Baumberger
found that the interfacial stiffness was proportional
to $F$ for nonadhesive solids with very different elastic properties~\cite{Berthoud98}.
The proportionality can be expressed as
\begin{equation}
K = p/u_0,
\label{eq:stiffness}
\end{equation}
where $K$ is the interfacial stiffness normalized by $A_0$,
and $p \equiv F/A_0$.
The characteristic length $u_0$ was found to be of order the combined root
mean squared (rms) roughness $h_\text{rms}$ ($\sim 1 \mu {\rm m}$) of the
surfaces.
The surfaces had self-affine fractal roughness that
is common in experiments.
Berthoud and Baumberger rationalized their observations within the contact
mechanics theory of Greenwood and Williamson~\cite{Greenwood66}, which,
however, is based on hypotheses that later turned out to
be unjustified~\cite{Persson07,Carbone08,Ramisetti11}.
Nonetheless, the results of additional
experiments~\cite{Benz:2006,Lorenz:2009} and
computer simulations of elastic contacts~\cite{Pei:2005,Benz:2006,Yang:2008,Carbone09,Campana10,Robbins11,Almqvist11,Carbone11}
with self-affine, fractal roughness, are consistent with
Eq. (\ref{eq:stiffness}).
Moreover, the proportionality coefficient $u_0$
agrees, to within ${\cal O}(10\%)$, with $u_0 \approx 0.4 h_\text{rms}$,
derived from the parameter-free contact mechanics theory of
Persson~\cite{Campana10,Almqvist11,Robbins11}.

The interfacial stiffness can be determined from the total stiffness
$K_{\rm tot}$ and the stiffnesses of ideal flat bounding solids
$K_1$ and $K_2$ using the rule for springs in series
\begin{equation}
{K^{-1}} \equiv K_{\rm tot}^{-1} - K_1^{-1} - K_2^{-1} \ .
\label{eq:indirectK}
\end{equation}
An alternative approach is to measure --- or to compute ---
the mean interfacial separation $\bar{u}$.
Changes in $\bar{u}$ are a direct measure of the deformation
attributable to the interface and
$K = - dp/d\bar{u}$
where the sign reflects the fact that $\bar{u}$ decreases with increasing
confining force.
In the range of validity of (\ref{eq:stiffness}),
this differential relation can be solved to yield
another testable prediction
\begin{equation}
p = p_0 {\rm exp} (-\bar u/u_0), \label{eq:oldEqno1}
\end{equation}
where $p_0$ is an integration constant.
Persson theory finds that $p_0 = \beta E^*$, where $E^*$ is the
effective elastic modulus and $\beta$ is dimensionless.
Like $u_0$, $\beta$ only depends on the spectral properties
of the surface~\cite{Persson07}.
Analytical expressions for $u_0$ and $\beta$ and computer simulations
agree again to within
${\cal O}(10\%)$~\cite{Robbins11,Almqvist11,Carbone11,Carbone09,Campana10}.

In a recent letter, Pohrt and Popov~\cite{Pohrt12} challenged the established
results on interfacial stiffness by
proposing a sublinear $K \propto p^\alpha$ power law deduced from numerical
simulations of an indenter with a square punch geometry.
Specifically, they reported $\alpha = 0.2567\times (3-H)$, where $H$ is the
Hurst roughness exponent.
This estimate was later corrected to $\alpha = 0.266 \times (3-H)$ and
scaling arguments were presented for a third relation $\alpha = 1/(1+H)$~\cite{Pohrt12PRE}.
Pohrt {\it et al.} argued that their results differed from previous ones because
their surfaces were ``{\it truly fractal}''~\cite{Pohrt12PRE},
i.e., roughness lived on wavelengths all the way to the
linear size $L_{\rm p}$ of their punch.
In particular they state: ``Whenever the surfaces are truly
fractal with no cut-off wavelength, a power law applies''~\cite{Pohrt12PRE}.

In this paper, we unravel the origin of the discrepancy
between the established results and the new findings.
To do so, we analyze finite-size effects in numerical simulations.
We derive analytical expressions, free of adjustable parameters,
that capture finite-size effects and constitute a complete theory for the stiffness of rough contacts.
For brevity, we present only the essence of the calculations in the main part
of this work.
Details on the numerical procedure can be found in Appendix~\ref{sec:num}.
The full derivation of prefactors for our scaling theory, can
be found in Appendix~\ref{sec:scaling}.
Appendix~\ref{sec:exp} contains unpublished experiments in support of Persson's contact mechanics theory.

\section{Numerical results}

We first summarize the arguments for how Eq. (\ref{eq:stiffness})
arises from the self-affinity of interfaces~\cite{Persson07}.
The key idea is that when there are a large number of separated contacting
patches, the distribution of contacts
is self-similar.
As the load increases, existing contact patches grow and new, small contacts
are formed.
This happens in such a way that the distributions of contact sizes and
local pressures remain approximately constant
over a wide range of loads~\cite{hyun04,Campana07}.
An immediate consequence is a linear relation between real contact area $A$
and $p$, which has been confirmed in many simulations, including all numerical
studies cited here.
The spatial correlations between contacting areas and local stresses
are also the same up to a prefactor that grows linearly with load
because of a sum rule \cite{Campana08}.
Since the system responds linearly, the elastic energy $U_\text{el}$ is
given by an integral of an elastic Greens function times
the Fourier transform of the stress-stress correlation function
and must thus be proportional to load:
\begin{equation}
\label{eq:elastEnerg}
U_\text{el} = u_0 A_0 p .
\end{equation}
Since the elastic energy
is equal to the work done by the external load
(assuming hard-wall interactions and no adhesion), it follows that
$dU_\text{el} = u_0 A_0 dp \equiv - A_0 p(\bar u) d \bar u$.
This last relation is identical to
(\ref{eq:oldEqno1}) and thus also to (\ref{eq:stiffness}).

When $p$ is so small that two {\it finite} surfaces start touching,
the interface cannot yet behave in a self-similar fashion.
The reason is that contact occurs
only near the highest asperity whose height
determines the separation at first contact $u_c$.
As a consequence, the validity of the arguments leading to
(\ref{eq:elastEnerg}) and
thus to (\ref{eq:stiffness})
--- or any theory valid in the thermodynamic limit ---
cannot hold at small $p$.
As already pointed out earlier, finite-size effects then become
important~\cite{Lorenz}.
Specifically, for a finite system $p$ vanishes for (finite) $\bar{u} > u_{\rm c}$,
while for an infinite system $p$ is always non-vanishing.
Thus, $p$ must initially decay faster with increasing $\bar u$  in a finite
system than in an infinite system where Eq.
(\ref{eq:stiffness}) holds.
In the opposite case of large $p$, a finite system approaches complete contact,
$\bar u = 0$, at finite pressure but infinite systems do not
because they have infinitely deep valleys.
One may conclude that contact formation of the highest peak and the lowest valley
depend on the specific realization of a surface.
However, for intermediate pressures, universal behavior may be found as long as
the roughness has well-defined statistical properties.

To study finite-size effects, we performed large-scale numerical
simulations of nonadhesive contact between a rigid self-affine surface
and an isotropic elastic substrate with effective modulus $E^*$ and
Poisson number $0.5$ using well-established methods
\cite{Campana:2006,Pastewka:2012} that are discussed in more detail in Appendix~\ref{sec:num}.
Surfaces were self-affine with Hurst exponent $H$ between
a short wavelength cut-off $\lambda_1$ and
long-wavelength roll-off $\lambda_r$
(see Fig.~\ref{fig:power_spectrum}).
The amplitudes of the Fourier transforms for the height
$\tilde{h}({\bf q})$ were drawn from a Gaussian distribution.
Their variances reflect the roughness spectrum $C(q)$ for  each reciprocal
space vector ${\bf q}$:
\begin{equation}
C(q) = C_0 \left\{
\begin{array}{ll}
0 & {\rm for} \ q < q_0
\\
1 & {\rm for} \ q_0 < q < q_r
\\
(q/q_r)^{-2-2H} & {\rm for} \ q_r < q < q_1
\\
0 & {\rm for} \ q_1 < q
\end{array}\right.
\label{eq:powerspectrum}
\end{equation}
Here, $q_0 = 2 \pi /L$, $q_1=2 \pi/\lambda_1$ and $q_r = 2 \pi/\lambda_r$ and the desired self-affine scaling is reflected in the power law for the range $q_r<q<q_1$.

\begin{figure}
\includegraphics[width=0.95\columnwidth]{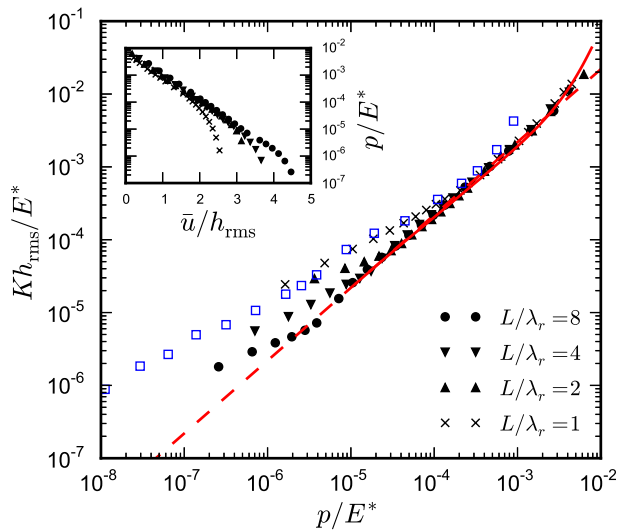}
\caption{\label{fig:test}
(Color online)
Log-log plot of the nondimensional contact stiffness $Kh_\text{rms}/E^*$ vs.
nondimensional pressure $p/E^*$ for self-affine
fractal surfaces with $H=0.7$ and rms slope $h_\text{rms}'=0.1$.
In all cases the  surface is resolved with $8192$ points in
each direction, $L/\lambda_1=4096$ and the ratio
of system size to roll-off wavelength, $L/\lambda_r$ is indicated.
The (red) solid line is the prediction of Persson's theory
while the dashed (red) line is the linear regime.
Open squares (blue) show the interfacial stiffness obtained from a punch calculation with $L_{\rm p}/\lambda_1=1024$.
Inset: Nondimensional pressure vs separation for the same surfaces.
}
\end{figure}

Figure~\ref{fig:test} shows typical results for the contact stiffness versus pressure.
Note that all the quantities are made dimensionless by dividing by the modulus and rms roughness so that they can be mapped to any experimental system with the same surface statistics.
In all cases, there is a linear relation at intermediate loads 
and a more rapid rise of $K$ with $p$ as full contact is approached.
Both regimes are well-described by Persson's contact mechanics theory (red line),
which requires only the surface roughness power spectrum and the
effective modulus as input.
We also find a transition to power law scaling at low loads.
This transition is
particularly sensitive
to the magnitude of a few random Fourier components
at the smallest wavevectors as well as to their relative phases.
The separation at first contact $u_c$ is also very sensitive to these Fourier components and
decreases with $L/\lambda_r$.
It cuts off the exponential relation between $p$ and $u$ shown in the inset.

Even for the case where $L/\lambda_r=1$, the results in
Figure~\ref{fig:test} follow linear scaling (Eq. (\ref{eq:stiffness}))
for more than one decade.
The range of validity of the linear scaling regime extends rapidly to lower $p$ as $L/\lambda_r$ increases.
Thus the more closely the thermodynamic limit is approached --- or the more
significant the statistical distribution of contacting peaks ---
the more accurate is Eq. (\ref{eq:stiffness}).
Given  typical $\lambda_r$,
e.g., ${\cal O}(10\mu{\rm m})$ for polished steel and ${\cal O}(1{\rm cm})$ for asphalt,
one can see that power law scaling matters only if $L/\lambda_r \approx 1$ or when loads are small.
Additionally, at extremely small loads where first asperities are touching the behavior should be Hertzian with $K \propto p^{1/3}$ if $\lambda_1>a_0$.
Indeed, we find that reducing the ratio of $\lambda_r/\lambda_1$ gives an exponent that approaches the one expected from Hertz contact mechanics.
Earlier reports of Hertzian-like behavior in load versus area for $\lambda_1>a_0$~\cite{Campana08b} are consistent with this finding.
This additional scaling regime limits the range where power-law scaling should be observed and complicates its measurement in simulations.

The results of Ref.~\cite{Pohrt12} do not show any appreciable region of linear scaling.
We have repeated their calculations to determine the origin of this discrepancy.
Instead of the periodic boundary conditions used here,
they considered a rigid, square punch with edge $L_{\rm p}$
contacting an infinite elastic substrate.
The punch had fractal roughness on all wavelengths from $\lambda_1$ to $L_{\rm p}$.
The open squares (blue) in Fig.~\ref{fig:test} show results for this
geometry (see also Appendix~\ref{sec:num}).
%
%
The interfacial stiffness was extracted from Eq.~\eqref{eq:indirectK} and the analytical punch solution~\cite{Pharr1992}.
While this correction is not performed in Ref.~\cite{Pohrt12}, it has little effect at the low loads of greatest interest.
%
%
%

At intermediate loads results for the flat punch and periodic boundary conditions follow the same power law scaling.
However, as in Ref.~\cite{Pohrt12}, the flat punch results cross over
to a rapid rise
with no region of linear scaling.
Inspection of the results shows that this behavior is associated with strong
artifacts from the boundary conditions at the edge of the punch.
The analytic solution for the pressure under a flat punch has a singularity
at the punch edges.
The solution for a rough punch approaches this solution as the pressure increases.
The pressure and stiffness are all dominated by regions near
the edge which approach full contact long before the central regions.
The strong influence of the edge makes the problem effectively one dimensional,
which may explain the success of the dimensional reduction used in Ref.~\cite{Pohrt12PRE} to fit their results.
%

\section{Scaling theory}

In the intermediate load regime, Fig.~\ref{fig:test} indicates
$K\propto p^\alpha$ behavior with $\alpha \approx 0.6$ for $H = 0.7$.
Thus, our small-pressure results for $L/\lambda_r=1$ are consistent with
Refs.~\cite{Pohrt12,Pohrt12PRE}.
In the following, we propose a new explanation for this power law by incorporating the
estimation of finite-size effects into Persson's contact mechanics theory.
The goal is to find an expression for the elastic energy
because it allows us to calculate the contact stiffness.
We reexpress a small change of the elastic energy
$dU_\text{el} = - p A_0 d\bar{u}$
as
$dU_\text{el} = - p A_0 dp (d\bar{u}/dp)$.
Inserting $K = -dp/d\bar{u}$ and $F = pA_0$ yields
\begin{equation}
p = K \frac{dU_\text{el}}{dF} .
\label{eq:oldEqno3}
\end{equation}

Our approach is motivated by the fact that the elastic energy is dominated by
the longest wavelength modes \cite{Campana10}.
For a single contacting region around the highest peak,
the longest wavelength will scale with the radius $r_0$
of the smallest circle that encloses the contacts.
We will first calculate the elastic energy $U_\text{el}^{(0)}$ for a
single Hertzian-like mesoscale asperity with radius of curvature $R$
and contact radius $r_0$.
Then we show that including roughness on the mesoasperity at wavelengths smaller than $r_0$
gives the same power law scaling for the elastic energy $U_\text{el}^{(1)}$.
For brevity, what follows presents only the general scaling arguments that explain the observed power law. A general derivation, including all prefactors, is given in Appendix~\ref{sec:scaling}.

An effective asperity radius is calculated from the
roughness at scales larger than $r_0$.
The local curvature $\nabla^2 h$ corresponds to
$q^2 h(q)$ in Fourier space.
Thus $R$ can be estimated as:
\begin{equation}
\frac {1}{R^2} \propto \int\limits_{q_0}^{\pi/r_0} d^2 q \ | q^2 h(q) |^2
\propto \int\limits_{q_0}^{\pi/r_0} dq \ q^5 C(q).
\label{eq:oldEqno4}
\end{equation}
For self-affine fractal roughness the surface roughness power spectrum is $C(q) \propto q^{-2-2H}$.
This gives $R \propto r_0^{2-H}$,
where we have assumed that the lower integration
bound to the last integral must be negligible at a small load.
This condition is fulfilled as long as $r_0 \ll \lambda_r$.

According to Hertzian contact mechanics, $r_0 \propto (RF)^{1/3}$.
Inserting $R \propto r_0^{2-H}$ and solving for $r_0$, we obtain
\begin{equation}
r_0 \propto F^{1/(1+H)}.
\end{equation}
The elastic energy stored within a Hertzian contact is $U_\text{el}^{(0)} \propto F \delta$ where the penetration depth $\delta \propto r_0^2/R \propto r_0^H$.
We obtain
\begin{equation}
U_\text{el}^{(0)} \propto F^{(1+2H)/(1+H)}
\label{Uel}
\end{equation}
and from Eq. (\ref{eq:oldEqno3})
\begin{equation}
K \propto p^{1/(1+H)}.
\label{powlaw}
\end{equation}

We now show that the elastic energy $U_\text{el}^{(1)}$
due to microscale roughness within the mesoscale asperity
also scales with $F^{(1+2H)/(1+H)}$.
The main assumption now is that the contact pressure within the mesoscale asperity
contact region is high enough that the contact mechanics theory by Persson
can be applied.
Then from (\ref{eq:elastEnerg}), $U_\text{el}^{(1)} = u_1 A_1 p_{\rm 1}$,
where $A_1=\pi r_0^2$ is the (nominal) contact area at the mesoscale
and $p_1 = F/A_1$.
The term $u_1$ is of order the rms roughness including only roughness
components with wavelength $\lambda < r_0$.
This can be written as
\begin{equation}
(h_\text{rms}^\text{meso})^2 = 2 \pi \int\limits_{\pi/r_0}^{2\pi/\lambda_1} dq \ q C(q)
\propto \left (\frac{\pi}{r_0}\right )^{-2H}-\left (\frac{2\pi}{\lambda_1}\right )^{-2H} .
\end{equation}
Since $\lambda_1 \ll r_0$ (unless $H$ is close to 0)
one obtains $h_\text{rms}^\text{meso} \propto r_0^{H}$
and $u_1 \propto r_0^H$.
Inserting $r_0 \propto F^{1/(1+H)}$, we get
$U_\text{el}^{(1)} \propto F^{(1+2H)/(1+H)}$
as in Eq.~(\ref{Uel}).

\begin{figure}[hbtp]
\includegraphics[width=0.95\columnwidth]{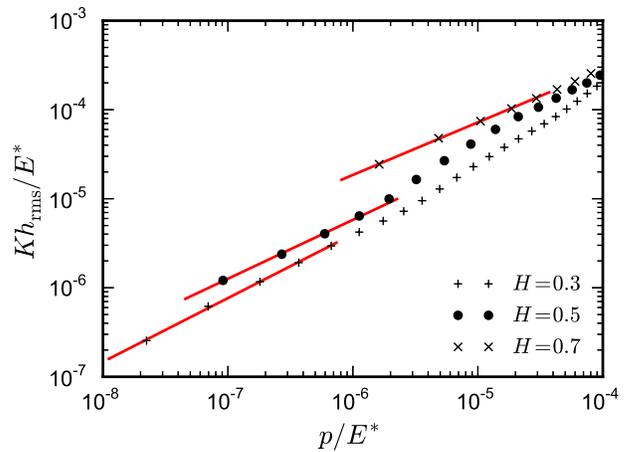}
\caption{\label{fig:scalH}
(Color online) Dimensionless interfacial stiffness $Kh_\text{rms}/E^*$ as a function of pressure $p/E^*$ in the finite-size region
for $L/\lambda_1=4096$ and different Hurst exponents $H$.  All calculations are for $L/\lambda_r=1$. Solid lines show a fit to the first four data points.
}
\end{figure}

 From the above treatment we predict that the stiffness $K$ scales
as $p^\alpha$ with $\alpha = 1/(1+H)$.
Fig.~\ref{fig:scalH} shows $K(p)$ relations obtained numerically
in the finite-size regime for different values of $H$.
Rough estimates for $\alpha$
were obtained by fitting to the lowest
four data points. The results from the simulations are:
$\alpha(H=0.3) = 0.72$ (see also below),
$\alpha(H=0.5) = 0.66$,  and
$\alpha(H=0.7) = 0.59$.
These values compare well to the theoretical predictions,
$\alpha(H=0.3) \approx 0.769$,
$\alpha(H=0.5) \approx 0.667$,  and
$\alpha(H=0.7) \approx 0.588$,
particularly
if one keeps in mind that
systematic simulation errors increase as $H$ approaches zero.

We note here that the expression for the microasperity contribution
to the total elastic energy depends on the elastic coupling between
the asperities. Any derivation neglecting this coupling~\cite{Pohrt12}
cannot describe the correct physics,
even if the resulting scaling is similar to $K\propto p^{1/(1+H)}$.
Moreover, probing the constitutive relation between pressure and
stiffness at a mesoscale will entail much larger fluctuations than
in a multi-asperity contact at the same pressure but larger
value of $L/\lambda_r$.

\begin{figure}
 \includegraphics[width=0.95\columnwidth]{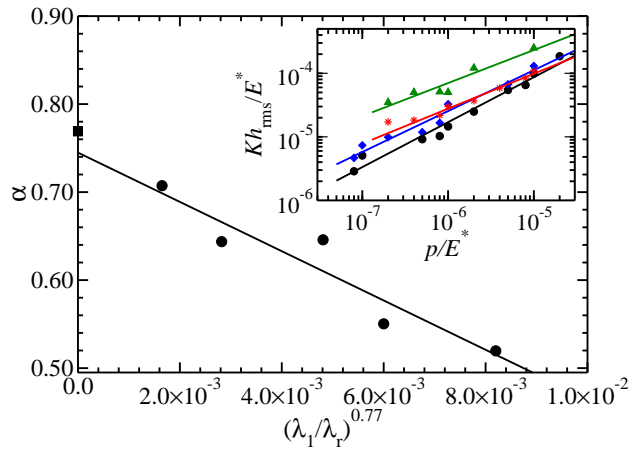}
 \caption{\label{fig:stiffness_ls}
(Color online) Exponent $\alpha$ as a function of $(\lambda_1/\lambda_r)^{0.77}$. Theoretically predicted value is denoted by $\blacksquare$.
The inset shows selected numerical data for $K(p)$ from which $\alpha$  was deduced. Data for the following values of $\lambda_r/\lambda_1$ is presented: 4096 ($\bullet$), 2048 ($\blacklozenge$), 768 ($\ast$), 512 ($\blacktriangle$).
 }
\end{figure}

The arguments that lead to $K\propto p^{1/(1+H)}$ hold when $\lambda_1\ll r_0\ll\lambda_r$.
Since $R\propto r_0^{2-H}$, the radius of the mesoasperity diverges as the contact area grows and $r_0\to\lambda_r$.
In this limit, the mesoasperities are flat, both
Eqns. (\ref{eq:indirectK}) and (\ref{eq:oldEqno1})
hold, and we rediscover the thermodynamic
limit $K\propto p$~\cite{Robbins11,Almqvist11,Carbone11,Carbone09,Campana10}.
On the other hand, if $r_0<\lambda_1$ the surface of the mesoasperity is smooth.
The upper integration bound in (\ref{eq:oldEqno4}) is then given by the short wavelength cut-off $q_1=2\pi/\lambda_1$ and $R$ is constant.
This ultimately must lead to traditional Hertz behavior where $K\propto p^{1/3}$.
Fig.~\ref{fig:stiffness_ls} shows the results of an attempted extrapolation to
the ``fractal limit'' $\lambda_1/\lambda_r \to 0$
for the value of $H = 0.3$, which had the largest discrepancy between theory and simulation.
Despite quite large stochastic scatter, we conclude that the
value of $\alpha = 1/(1+H)$ is consistent with the simulations.

Finally, we address how the finite size power law region depends on linear system size $L$ and roll-off length $\lambda_r$.
Following along the lines of the above derivation, it is straightforward to compute the full expression for the interfacial stiffness (see Appendix~\ref{sec:scaling}):
\begin{equation}
\label{eq:oldEqno5}
\frac{K h_\text{rms}}{E^*}
=
\theta
\left(
\frac{h_\text{rms}}{2\pi \lambda_r} \frac{\lambda_r^2}{L^2}
\right)^{H/(1+H)}
\left(
 \frac{p}{s^{1/2} E^*}
 \right)^{{1}/{(1+H)}},
\end{equation}
where $1/s=1+H[1-(\lambda_r/L)^2]$.
The prefactor $\theta$ depends only on the Hurst exponent $H$, but for $H>0.3$ variation is restricted to $0.75\lesssim\theta\lesssim1.0$ (see also Fig.~\ref{1H.2invThete1.invTheta2.invSum}).
By equating (\ref{eq:oldEqno5}) with $K=p/\gamma h_\text{rms}$, where $\gamma \approx 0.4$, we obtain an estimate for the pressure $p_c$ at which the stiffness crosses over from power law to linear behavior:
\begin{equation}
\label{eq:oldEqno6}
\frac{p_c}{E^*}
=
\frac{h_\text{rms}}{2\pi \lambda_r}
\frac{\lambda_r^2}{L^2}
s^{-1/2H}
\left(\theta \gamma \right)^{{(1+H)}/{H} }.
\end{equation}
For different realizations of the surface the prefactor of the power law and $p_c$ can vary significantly.
Nevertheless, for the data shown in Fig.~\ref{fig:test} we find $p_c/E^* \approx 6\times 10^{-5}$ for $q_r/q_0=1$ and
$p_c/E^* \approx 3\times 10^{-6}$ for $q_r/q_0=8$, in excellent agreement with the numerical data.
Generally, the cross-over pressure $p_{\rm c}$ decreases with increasing linear system size $L$.
Equation~(\ref{eq:oldEqno6}) also reveals the importance of separation between $L$ and the roll-off length $\lambda_r$.
Scale separation pushes the crossover to lower pressure even more rapidly since the ratio $L/\lambda_r$ enters quadratically.
In the thermodynamic limit $L/\lambda_r\to\infty$, the power law region vanishes all together.

\section{Conclusions}

We conclude that the previously reported
$K \propto p$  and $p \propto \exp(-\bar u/u_0)$ laws \cite{Benz:2006,Lorenz:2009,Pei:2005,Yang:2008,Carbone09,Campana10,Robbins11,Almqvist11,Carbone11} are satisfied when there is a statistical ensemble of high peaks in contact.
This linear scaling extends to lower loads as the upper length scale of roughness
decreases, because there is a better statistical sampling of high peaks.
At the smallest loads, the contact diameter is smaller than the
smallest wavelength of roughness, and the stiffness follows
the Hertz expression for contact of a single spherical asperity and $K\propto p^{1/3}$.
At intermediate contact areas and loads, contact is confined to a single
large peak with a fractal hierarchy of smaller bumps.
In this regime $K$ scales sublinearly with $p$ and the prefactor and
corresponding surface separation
have large fluctuations from one sample to the next even in the limit of large system size.
Parameter-free expressions for the power law $\alpha=1/(1+H)$ and prefactor
(Eq.~\eqref{eq:oldEqno5}) were derived.
The power law agrees with one of the results presented in Ref.~\cite{Pohrt12PRE}, although they also presented linear $\alpha \propto H$ expressions~\cite{Pohrt12,Pohrt12PRE} when fitting their numerical data.
Ref.~\cite{Pohrt12PRE} also discussed the scaling of the prefactor with $L$
but we provide a full expression including the dependence on rolloff $\lambda_r$
and Hurst exponent $H$.
 
Pohrt, Popov and Filipov (Refs.~\cite{Pohrt12} and \cite{Pohrt12PRE})
found no linear regime in their studies of stiffness.
In part this was because they considered the limiting case of roughness at wavelengths up to the size of their contact ($\lambda_r=L_{\rm p}$).
As recently pointed out by Barber~\cite{Barber13}, statistical fluctuations make a prediction of stiffness (and related properties such as conductance) difficult if there is no separation between the scales of the macroscopic object and the longest wavelength of the roughness.
For nominally flat surfaces and periodic boundary conditions, we observe that the linear $K\propto p$ regime holds for at least an order of magnitude in load even in this extreme case.
The square punch geometry considered in Ref.~\cite{Pohrt12} suppresses
this linear regime (Fig.~\ref{fig:test}).
Stress is concentrated near the edges of the punch, which approach full
contact long before the central region.
This pronounced heterogeneity makes the punch geometry a poor choice and
it is rarely used in experiments because of the difficulty in achieving
perfect alignment~\cite{Pelletier}.

Most experimental realizations of surfaces have an rms roughness and 
upper cut-off on fractal scaling that are both
significantly smaller than the system size.
As a result, Eq.~\eqref{eq:oldEqno6} predicts that
the linear relation between stiffness and load should extend
over the experimental range.
Indeed, measurements by Berthoud and Baumberger~\cite{Berthoud98} show
$K\propto p$ at fractional contact areas of $10^{-6}$ and below.
We conclude that as long as the contact responds elastically,
the power law region appears to be confined to low pressure that is difficult to access in macroscopic experiments, and has therefore little impact on most applications.

As an example consider applications to syringes, where the relation between the squeezing pressure $p$ and the average
interfacial separation $\bar u$ (which determines the contact stiffness) is very important for the fluid leakage at the
rubber-stopper barrel interface~\cite{Dapp}. The key contact region is between a rib of the rubber stopper and the barrel.
The width of the contact region (of order $w\approx 1 \ {\rm mm}$) defines the cut-off wavevector $q_r = 2\pi /w \approx 6000 \ {\rm m}^{-1}$.
The Hurst exponent $H\approx 0.9$ and the rms roughness amplitude
(including the roughness components with wavevector $q > q_r$)
is $h_{\rm rms} \approx 3 \ {\rm \mu m}$. The elastic modulus of the rubber stopper is typically $E\approx 3 \ {\rm MPa}$.
Using these parameters we find from Eq. 13 that the stiffness should rise linearly with pressure above $p_{\rm c} \approx 1 \ {\rm kPa}$.
This is negligible compared to the pressure in the contact region between the rib of the rubber stopper and the barrel,
which is typically of order $\sim 1 \ {\rm MPa}$.
 
As devices shrink towards the nanoscale, $h_{\rm rms}$ and $\lambda_{\rm r}$
may become closer to the system size.
%
For example, Buzio et al.~\cite{Buzio} report nonlinear stiffness when loading
flat contacts of size $L_{\rm p} \sim 2\, {\rm \mu m}$ on rough surfaces with $h_{\rm rms} \sim 20\, {\rm nm}$ to $100\,{\rm nm}$ and $\lambda_L \sim 1\, {\rm \mu m}$ up to forces of $200\,{\rm nN}$.
Eq.~\eqref{eq:oldEqno6} predicts nonlinear behavior for these parameters,
but the experimental tips were adhesive,
there was evidence of plastic deformation, and atomistic effects
may become important at nanometer scales \cite{luan05}.
None of these effects has been included here or in Refs.~\cite{Pohrt12} and \cite{Pohrt12PRE} and future work on their influence will be of great interest

\acknowledgments
This material is based upon work supported by the U.S. National Science Foundation under Grants No. DMR-1006805, OCI-0963185 and CMMI-0923018.
NP and MHM thank the J\"ulich Supercomputing Centre for computing time. MHM thanks DFG for financial support through grant Mu 1694/5-1. MR acknowledges a Simons Foundation Fellowship
and LP acknowledges support from
the European Commission under Marie-Curie IOF-272619.

\appendix

\section{Details of the numerical calculations}\label{sec:num}

\begin{figure}
\includegraphics[width=0.95\columnwidth]{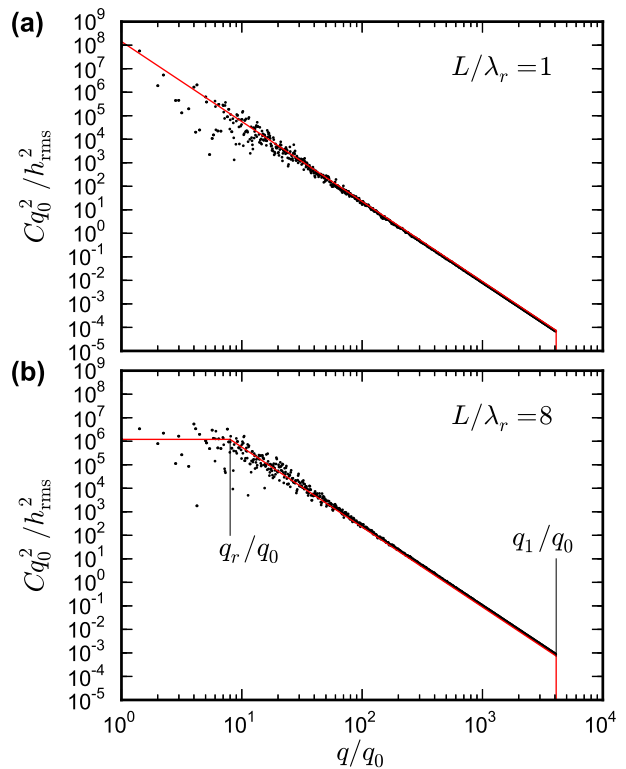}
\caption{\label{fig:power_spectrum}
(Color online) Power spectra for two surfaces without (a) and with (b) a roll-off at large wavelength as generated by a Fourier filtering algorithm. The solid lines show the prescribed power spectrum $C(q)$ and the dots the actual realization. Panel (b) indicates the wavevectors of the long-wavelength roll-off $q_r=2\pi/\lambda_r$ and the short-wavelength cut-off $q_1=2\pi/\lambda_1$. For $q<q_0=2\pi/L$ where $L$ is the linear system size the surfaces have zero power. The noise at low $q$ is due to the fact that order $q^2$ Fourier components contribute to the power-spectrum at wavevector $q$ of a realization of a surface.
}
\end{figure}

Self-affine rough surfaces with the desired $H$, $h_{\rm rms}'$, $\lambda_s$ and
$\lambda_L$ were generated using a Fourier-filtering algorithm described
previously~\cite{Ramisetti11}.
Fourier components for each wavevector ${\bf q}$ have a random phase and a normally distributed amplitude that depends on the wavevector magnitude $q$ according to Eq.~\eqref{eq:powerspectrum}.
Periodic boundary conditions with period $L$ were applied in the plane of the surface to prevent edge effects.
Fig.~\ref{fig:power_spectrum} shows a roughness power spectrum as generated by this algorithm and used in the simulations.
The solid lines indicate the mean values for the spectrum, while the dots
reflect one particular realization.
Fluctuations of the height $h({\bf r})$ in real space are not only the consequence of variations in the absolute value of their complex Fourier transforms $\tilde{h}({\bf q})$ but also due to the random phases.
From Fig.~\ref{fig:power_spectrum} it becomes clear that the largest fluctuations occur at small wavevectors (large wavelength) because $q^2$ Fourier components contribute to a realization at wavevector ${\bf q}$.

We considered elastic substrates with contact modulus $E^*$ and Poisson ratio $\nu=1/2$.
At $\nu=1/2$ the in-plane components and the out-of-plane components of the elastic displacement field decouple.
We then only treated the out-of-plane components $u({\bf r})$ on a grid with spacing $a_0$.
More specifically, we carried out simulations with $E^*=2$ and $a_0=1$, but since all quantities are presented here in a dimensionless form the actual values of these quantities do not matter.
The elastic interaction was solved using a Fourier-transform technique~\cite{Campana:2006,Pastewka:2012} that accelerates computation of the force $f({\bf r})=\int d^2r' G^{-1}({\bf r}-{\bf r}') u({\bf r}')$.
For periodic calculations, we used a linearized surface Green's function~\cite{Pastewka:2012,AmbaRao}.
In reciprocal space, the expression for the Green's function is $\tilde{G}^{-1}({\bf q})= E^* q/2$.
For nonperiodic calculations, we employed a real-space surface Green's function $G({\bf r})$ that is derived from the elastic response to a uniform pressure on a square region of area $a_0^2$~\cite{Johnson1985}.
A padding region was used to separate repeating images~\cite{Hockney}.

\section{Full scaling theory, including derivation of all prefactors}\label{sec:scaling}

Consider a randomly rough surface with a power spectrum given by Eq.~\eqref{eq:powerspectrum} and shown in Fig. \ref{fig:power_spectrum}.
The surface mean square roughness amplitude is then given by
\begin{align}
h_{\rm rms}^2 &= \int d^2 q \ C(q) \\
  &= 2\pi C_0 \left [\int\limits_{q_0}^{q_r}
dq \ q+ \int\limits_{q_r}^{q_1} dq \ q \left ({q\over q_r}\right )^{-2-2H}\right ] \\
  &\approx {\pi q_r^2 \over Hs} C_0,
\end{align}
where $1/s=1+H\left [1-(q_0/q_r)^2\right ]$ and the last equality holds in the limit $q_1/q_r \gg 1$.
Note that $s=1$ if $q_0=q_r$ and there is roll-off region.
Expressed in terms of $h_{\rm rms}$, the normalization of the power-spectrum is
\begin{equation}
C_0 = {Hs\over \pi q_r^2} h_{\rm rms}^2.
\end{equation}


\subsection{Hertzian-like mesoscale asperity}

We first calculate the elastic energy stored in the deformation field associated with the Hertz mesoscale asperity contact region.
The mesoscale asperity has the radius of curvature $R$.
The radius of the (apparent) contact region between the mesoscale asperity and the
flat countersurface is denoted by $r_0$.
We assume that no roughness lives on scales $<r_0$ such that the mesoscale asperity can be treated as smooth.
The mean summit asperity curvature is given by~\cite{Nayak}
$\bar \kappa = \beta \surd 2 \kappa_0$
where $\kappa_0$ is the root-mean-square curvature of the surface:
\begin{equation}
\kappa^2_0 = {1\over 2} \int d^2q \ q^4 C(q) = \pi \int\limits_{q_0}^{q_1} dq \ q^5 C(q).\end{equation}
Nayak~\cite{Nayak} has shown that $\beta = \sqrt{8/3\pi}$ when roughness occurs on many length scales so that $q_1/q_0 \gg 1$.
If we include only roughness components with wavevector $q < \pi /r_0$, then we obtain the mean summit curvature $1/R$ of the mesoscale asperity:
\begin{equation}
{1\over R^2} = 2\pi \beta^2  \int\limits_{q_0}^{\pi /r_0} dq \ q^5 C(q).
\end{equation}
We now define the dimensionless quantities $\bar R = q_rR$, $\bar h_{\rm rms} = q_r h_{\rm rms}$ and $\bar r_0 = q_r r_0$. This gives the mean dimensionless summit curvature
\begin{align}
{1\over \bar R^2} &= \frac{2\pi\beta^2 C_0}{q_r^2}\left [\int\limits_{q_0}^{q_r} dq \ q^5 +\int\limits_{q_r}^{\pi/r_0} dq \ q^5 \left ({q\over q_r}\right )^{-2-2H}\right ] \\
&\approx
{Hs \beta^2\over 2-H} \bar h_{\rm rms}^2 \left ({\pi \over \bar r_0} \right )^{4-2H},
\label{eq:recR}
\end{align}
where the last equality holds in the limit $\pi /r_0 \gg q_r$.
We define the dimensionless prefactor
\begin{equation}
\chi = \left (\frac{2-H}{Hs}\right )^{1/2} \frac{\pi^{H-2}}{\beta\bar{h}_{\rm rms}}.
\label{eq:alpha}
\end{equation}
and simplify Eq.~\eqref{eq:recR} to
\begin{equation}
  \bar{R} = \chi \bar r_0^{2-H}.
  \label{eq:R}
\end{equation}

We now use Hertz theory to obtain the mesoasperity radius $r_0$ as a function of normal force $F$.
Hertz theory gives a dimensionless mesoasperity contact radius of
\begin{equation}
\bar r_0^3 = {3\over 4} \bar F \bar R = {3 \chi \over 4} \bar F \bar r_0^{2-H},\end{equation}
where $\bar F = Fq_r^2/E^*$ is the dimensionless normal force and we used Eq.~\eqref{eq:R} for the dimensionless asperity radius.
We now solve for $\bar{r}_0$ to obtain:
\begin{equation}
\bar r_0 = \left ( {3 \chi \over 4} \bar{F} \right )^{1/(1+H)}.
\label{eq:r0}
\end{equation}
By inserting this expression into Eq.~\eqref{eq:R} the mesoasperity radius becomes
\begin{equation}
\bar R = \chi^{3/(1+H)} \left (\frac{3}{4} \bar{F}\right )^{(2-H)/(1+H)}.
\label{eq:Rfinal}
\end{equation}

The elastic energy stored in the Hertz mesoscale deformation field for depth of indentation $\delta$ is given by
\begin{equation}
\bar U_{\rm el}^{(0)} = {2\over 5} \bar F \bar \delta,
\label{eq:Uel0first}
\end{equation}
where $\bar \delta = q_r \delta$ and the dimensionless energy $\bar{U}^{(0)}_{\rm el}=U_{\rm el}^{(0)}q_r^3/E^*$.
Hertz theory also tells us the displacements as a function of normal force
\begin{equation}
\bar \delta = \left ( {9\bar F^2\over 16 \bar R}\right )^{1/3}
 = \chi^{-1/(1+H)} \left(\frac{3}{4} \bar{F}\right)^{H/(1+H)},
\label{eq:delta}
\end{equation}
where we used Eq.~\eqref{eq:Rfinal} to substitute the mesoasperity radius.
By combining Eqs.~\eqref{eq:Uel0first} and \eqref{eq:delta} the elastic energy becomes
\begin{equation}
\bar U_{\rm el}^{(0)}
= \frac{4}{3} \kappa_0 \chi^{-1/(1+H)} \left (\frac{3}{4} \bar{F}\right )^{(1+2H)/(1+H)}
\label{eq:Uel0final}
\end{equation}
with $\kappa_0=2/5$.

\subsection{Microscale roughness on mesoscale asperity}

Next we calculate the elastic deformation energy that is stored in microasperity contacts within the Hertz mesoasperity contact region~\cite{Persson07,Campana10}.
This energy is given by Eq.~\eqref{eq:elastEnerg}:
\begin{equation}
U_{\rm el}^{(1)} = u_1 A p_1 = u_1 F.
\label{eq:Uel1}
\end{equation}
In terms of the dimensionless quantities it becomes $\bar U_{\rm el}^{(1)} = \bar u_1 \bar F$ where $\bar u_1 = q_r u_1$.
Additionally, we have $\bar u_1 = \gamma (\bar{h}_{\rm rms}^{\rm meso})$ where $\gamma \approx 0.4$.
Note that $h_{\rm rms}^{\rm meso}$ is the root mean square roughness amplitude within the mesoasperity, i.e. within the area confined by the mesoasperity contact radius $r_0$.
In contrast $h_{\rm rms}$ is the root mean square roughness amplitude of the full surface all the way to the linear system size $L$.
We can express $h_{\rm rms}^{\rm meso}$ in terms of $h_{\rm rms}$:
\begin{align}
(h_{\rm rms}^{\rm meso})^2 &=
2\pi C_0 \int\limits_{\pi/r_0}^{q_1} dq \ q \left ({q\over q_r}\right )^{-2-2H}
\\
&\approx  s h_{\rm rms}^2 \left( \frac{\pi/r_0}{q_r} \right )^{-2H}.
\end{align}
This holds for $\pi/r_0\ll q_1$.
In terms of the dimensionless quantities this becomes:
\begin{equation}
\bar{h}_{\rm rms}^{\rm meso} = s^{1/2} \bar h_{\rm rms} \left(\frac{\pi}{\bar{r}_0} \right)^{-H}.
\end{equation}
We now use the definition for $\chi$ Eq.~\eqref{eq:alpha} to eliminate $ \bar h_{\rm rms}$ and use Eq.~\eqref{eq:r0} to express $\bar{r}_0$ in terms of the force $\bar{F}$:
\begin{equation}
\bar{h}_{\rm rms}^{\rm meso}
=
\left (\frac{2-H}{\pi^4\beta^2 H}\right )^{1/2} \chi^{-1/(1+H)} \left (\frac{3}{4} \bar{F}\right )^{H/(1+H)}.
\label{eq:hrms}
\end{equation}
By combining Eqs.~\eqref{eq:Uel1} and \eqref{eq:hrms} the elastic energy becomes
\begin{equation}
\bar U_{\rm el}^{(1)} =
\frac{4}{3} \kappa_1 \chi^{-1/(1+H)} \left (\frac{3}{4} \bar{F}\right )^{(1+2H)/(1+H)}
\label{eq:Uel1final}
\end{equation}
with
\begin{equation}
  \kappa_1=\gamma \left ( \frac{2-H}{\pi^4 \beta^2 H}\right )^{1/2}.
\end{equation}
Note that the expression for $\bar{U}_{\rm el}^{(1)}$ has the same form as the expression derived for the Hertz-like mesoasperity contact Eq.~\eqref{eq:Uel0final}.
They differ only in the prefactors $\kappa_0$ and $\kappa_1$.

\subsection{Total elastic energy and stiffness}

The total elastic energy is now given by the sum of the two contributions Eqs.~\eqref{eq:Uel0final} and \eqref{eq:Uel1final}, i.e. $\bar U_{\rm el} = \bar U^{(0)}_{\rm el}+\bar U^{(1)}_{\rm el}$.
This yields
\begin{equation}
\bar U_{\rm el}
=
\frac{4}{3} \kappa
 \chi^{-1/(1+H)}
 \left(\frac{3}{4}\bar{F}\right)^{(1+2H)/(1+H)}
 \label{eq:Ueltot}
\end{equation}
with $\kappa=\kappa_0+\kappa_1$.
We now compute the total dimensionless stiffness $\bar{k}=q_r K A_0/E^*$ from Eq.~\eqref{eq:oldEqno3}.
It is given by
\begin{equation}
\bar{k} = {\bar F \over d\bar U_{\rm el}/d\bar F},
\end{equation}
and inserting Eq.~\eqref{eq:Ueltot} yields
\begin{equation}
\bar{k} =  \theta \left (\frac{\bar{F}}{\bar{h}_{\rm rms} s^{1/2}}\right )^{1/(1+H)}.
\label{eq1}
\end{equation}
Reintroducing the dimensional quantities yields Eq.~\eqref{eq:oldEqno5}.

\begin{figure}[tbp]
\includegraphics[width=0.95\columnwidth,angle=0]{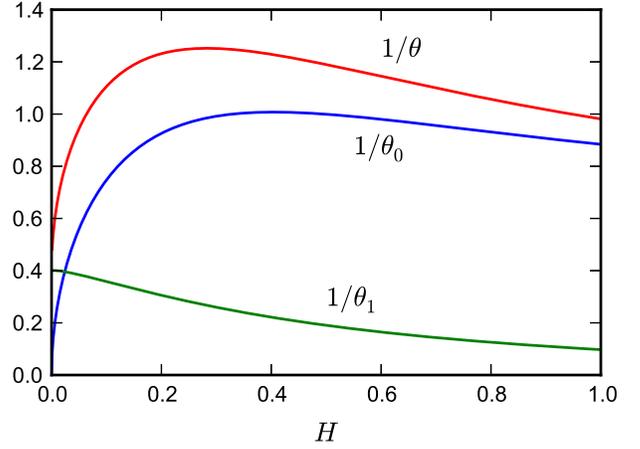}
\caption{(Color online) Plot of the values of $1/\theta_0$, $1/\theta_1$ and $1/\theta= 1/\theta_0+1/\theta_1$ as a function of Hurst exponent $H$. The quantities $\theta_0$ and $\theta_1$ are defined in the text.}
\label{1H.2invThete1.invTheta2.invSum}
\end{figure}

The dimensionless prefactor $\theta$ is given by two contributions as $1/\theta=1/\theta_0+1/\theta_1$ that each depend on the Hurst exponent $H$ only:
\begin{equation}
{1\over \theta_{0/1}}
=
 {1+2H\over 1+H} \left ({3\over 4\pi}\right )^{H/(1+H)} \left ({\pi^4 \beta^2 H\over 2-H}\right )^{1/(2+2H)} \kappa_{0/1}.
\end{equation}
In Fig.~\ref{1H.2invThete1.invTheta2.invSum}
we show $1/ \theta_0$, $1/ \theta_1$ and $1/\theta$ as a function of the Hurst exponent $H$.
It is interesting to note that as $H\rightarrow 0$,
then $1/ \theta_0 \rightarrow 0$ while $1/ \theta_1$ remains finite, i.e., for the fractal dimension $D_{\rm f} = 3-H = 3$
the stiffness is entirely determined by the  short-wavelength roughness in the mesoasperity contact region.
Note also that since $q_r \approx \pi /L$, where $L$ is the linear size of the system,
the stiffness scales as $k\sim q_r^{-H/(1+H)} \sim L^{H/(1+H)}$ with the size of
the system. This is in contrast to the region where $p\sim {\rm exp}(-u/u_0)$.
There, the interfacial contact stiffness is independent of the
size $L$ of the system. Note also that the stiffness scales with the
rms roughness as $h_{\rm rms}^{-1/(1+H)}$ while in the region $p\sim {\rm exp}(-u/u_0)$
the stiffness is proportional to $h_{\rm rms}^{-1}$.
For the Hurst exponent $H\approx 0.8$, which is typical in practical applications, $\theta \approx 1$, which appears to be in
good agreement with the prefactor found by Pohrt and Popov in their numerical simulation study~\cite{Pohrt12}.
The treatment
presented above can be generalized to obtain the distribution of stiffness values (at least approximately)
by calculating the distribution $P(R)$ of summit curvature radius $R$.

It is interesting to determine the critical force $F_{\rm c}$ such that for $F < F_{\rm c}$ one needs to use the finite
size power-law expression for the stiffness while for $F > F_{\rm c}$ the Persson expression is valid.
When the relation $p\sim {\rm exp} (-\bar u/u_0)$ is valid the stiffness is given by Eq.~\eqref{eq:stiffness}:
\begin{equation}
\bar{k}={\bar{F} \over \bar{u}_0} = {\bar{F} \over \gamma \bar{h}_{\rm rms}}.
\label{eq2}
\end{equation}
The critical force $F_{\rm c}$ is determined by the condition that $\bar{k}$ given by Eqs.~\eqref{eq1} and~\eqref{eq2} coincide. This gives
\begin{equation}
\theta \left ({\bar{F}_{\rm c} \over \bar{h}_{\rm rms} s^{1/2}}\right )^{1/(1+H)}={\bar{F}_{\rm c} \over \gamma \bar{h}_{\rm rms}}\end{equation}
which we can solve for the dimensionless critical force $\bar{F}_c$:
\begin{equation}
\frac{\bar{F}_{\rm c}}{\bar{h}_{\rm rms}} = s^{-1/2H} \left (\theta \gamma \right )^{(1+H)/H}.
\label{eq3}
\end{equation}
Reintroducing dimensional quantities yields Eq.~\eqref{eq:oldEqno6}.
%

\subsection{Discussion}

The prediction Eq.~\eqref{eq:oldEqno6} for the switching between the finite size region and the region where the stiffness is proportional to the loading force is in good agreement with our
simulation results.
The surfaces we have studied in numerical simulations have rms slope $h_{\rm rms}^\prime=0.1$ and $q_0/q_1 = 1/4096$ and $H=0.7$.
For our particular realizations we find $q_r h_{\rm rms} \approx 5.7\times 10^{-3}$ for $q_r/q_0 =
L/\lambda_r =1$ and $q_r h_{\rm rms} \approx 1.3\times 10^{-2}$ for $q_r/q_0 = 8$.
With these number we get $p_{\rm c}/ E^* \approx 6\times 10^{-5}$ for $q_r/q_0 = 1$ and $p_{\rm c}/ E^* \approx 3\times 10^{-6}$ for $q_r/q_0 = 8$  from Eq.~\eqref{eq:oldEqno6},
which is in good agreement with Fig.~\ref{fig:test}.
For the surface with $H=0.3$ we obtain (for a surface with rms slope $0.1$)
$q_rh_{\rm rms}$ nearly $100$ times smaller than for $H=0.7$, which will shift
the cross-over force $F_{\rm c}$, between the two stiffness regions, with a similar factor to lower values, again in good agreement
with the numerical studies.
The results presented above differ from the conclusion of Pohrt and Popov who state that the power-law relation observed for small
applied forces is valid for all applied forces~\cite{Pohrt12,Pohrt12PRE}.
 The present study shows that this statement is incorrect and Fig. 1 clearly shows that the
contact stiffness cannot be described by a power law for all applied forces as this would correspond to a straight line on our
log-log scale.

\section{Experiments}\label{sec:exp}

The relation~\eqref{eq1} as well as the above mentioned finite-size effect region has also been observed in experiments. In these
experiments a rectangular block of silicon rubber (a nearly perfect elastic material even at large strain) is squeezed
against hard, randomly rough surfaces. In this case no plastic deformation will occur, and the compression
of the rectangular rubber block, $(p/E')d$ (see below), which will
contribute to the displacement $s$ of the upper surface of the block, can be accurately taken into account. Such measurements were performed
in Ref. \cite{Lorenz}, and were found to be in good agreement with the theory (these tests involved no fitting parameters as the surface roughness
power spectrum, and the elastic properties of the rubber block, were obtained in separate experiments).
Here we show the result for the contact stiffness $K=-dp/d\bar u$ (not presented in Ref. \cite{Lorenz}) of one additional such measurement.

\begin{figure}[thbp]
\includegraphics[width=0.95\columnwidth]{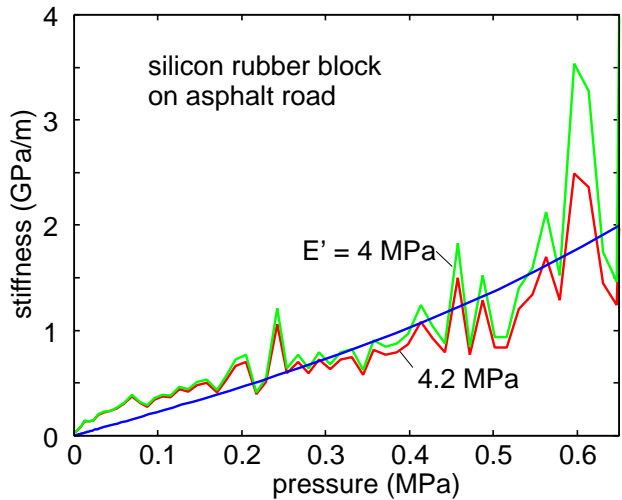}
\caption{\label{experiment}
(Color online) The normal contact stiffness as a function of the applied nominal contact pressure for
a silicon rubber block (cylinder shape with diameter $D=3 \ {\rm cm}$ and height $d=1 \ {\rm cm}$)
squeezed against a road asphalt surface. The green and red lines are obtained from the measured
$p(s)$ relation using (7) with $E'=4.0 \ {\rm MPa}$ and $4.2 \ {\rm MPa}$ (see text) while the blue line is the
theory prediction.
}
\end{figure}

The experiment was performed for a silicon rubber block (cylinder shape with diameter $D=3 \ {\rm cm}$ and height $d=1 \ {\rm cm}$)
squeezed against a road asphalt surface with the rms roughness amplitude $0.63 \ {\rm mm}$ and the roll-off wavelength $\lambda_L \approx
0.3 \ {\rm cm}$ as inferred from the surface roughness power spectrum.
The squeeze-force is applied via a flat steel plate and no-slip of the rubber
could be observed against the steel surface or the asphalt surface.
We measured the displacement $s$ of the upper surface of the block as a function of the
applied normal load. Note that
\begin{equation}
s=(u_{\rm c}-\bar u)+(p/E')d,
\label{eq6}
\end{equation}
where $E'$ is the effective Young's modulus taking into account the no-slip
boundary condition on the upper and lower surface, which was measured to be $E'=4.2 \ {\rm MPa}$ in a separate experiment where the
rubber block was squeezed between two flat steel surfaces. Using Eq.~\eqref{eq6} gives
\begin{equation}
K=-{dp \over d\bar u} = -{dp \over ds} {ds\over d\bar u} = {dp \over ds} \left (1+{K d \over E'}\right )\end{equation}
or
\begin{equation}
K={K^* \over 1-K^* d/E'},
\label{eq7}
\end{equation}
where $K^* = dp/ds$. Using (C3) we obtain the results shown in Fig. \ref{experiment}, which presents the normal contact stiffness as a function of the applied nominal contact pressure
obtained from the measured $p(s)$ relation with $E'=4.2 \ {\rm MPa}$ (measured value)
and $E'=4 \ {\rm MPa}$ (to indicate the sensitivity of the result to $E'$). For very small contact pressures $K^* \approx 0$
so that the denominator in (C3) is $\approx 1$ (and $K\approx K^*$ as assumed in Ref.
\cite{Pohrt12} without proof) and the result is insensitive to $E'$ as also seen in Fig. \ref{experiment}.
For large contact pressure the experimental data exhibits rather large noise (and great sensitivity to $E'$),
which originates from the increasing importance of the compression of the rubber block for large contact pressure. That is,
for large pressures the denominator in (C3) almost vanishes, which implies that a small uncertainty in the measured $p(s)$ relation
(which determines $K^*$), or in $E'$,  will result in a large uncertainty in $K$ for large pressures.

The blue curve in Fig. \ref{experiment} is the theory prediction which is obtained
without any fitting parameter using the measured surface roughness power spectrum. For small contact pressure the contact stiffness
obtained from the measured data is larger than predicted by the theory,
but for nominal contact pressures typically involved in rubber applications (which are $\sim 0.4 \ {\rm MPa}$ as in tire applications,
or higher in most other applications) the finite size effects are not important.


\begin{thebibliography}{99}

\bibitem{bowden56}
F. P. Bowden and D. Tabor,
{\it Friction and Lubrication}
(Wiley, New York, 1956).

\bibitem{dieterich94}
J. H. Dieterich and B. D. Kilgore,
Pure Appl. Geophys.  {\bf 143}, 283 (1994).

\bibitem{persson01}
B. N. J.  Persson,
J. Chem. Phys. {\bf 115}, 3840 (2001).

\bibitem{hyun04}
S. Hyun, L. Pei, J.-F. Molinari, and M. O. Robbins,
Phys. Rev. E {\bf 70},  026117 (2004).

\bibitem{Berthoud98}
P. Berthoud and T. Baumberger,
Proc. R. Soc. A {\bf 454}, 1615 (1998).

\bibitem{Barber03}
J. R. Barber,
Proc. R. Soc. A {\bf 459}, 53 (2003).

\bibitem{Greenwood66}
J. A. Greenwood and J. B. P. Williamson,
Proc. R. Soc. A {\bf 295}, 300 (1966).

\bibitem{Persson07}
B. N. J. Persson, Phys. Rev. Lett. {\bf 99}, 125502 (2007).

\bibitem{Carbone08}
G. Carbone and F. Bottiglione,
J. Mech. Phys. Solids {\bf 56}, 2555 (2008).

\bibitem{Ramisetti11}
S. B. Ramisetti, C. Campa\~{n}\'a, G. Anciaux, J.-F. Molinari,
M. H. M\"user, and M. O. Robbins,
J. Phys. Condens. Matter {\bf 23}, 215004 (2011).

\bibitem{Benz:2006}
M. Benz, K. J. Rosenberg, E. J. Kramer and J. N. Israelachvili,
J. Phys. Chem. B {\bf 110}, 11884 (2006).

\bibitem{Lorenz:2009}
B. Lorenz and B. N. J. Persson, J. Phys.: Condens. Matter {\bf 21}, 015003 (2009).

\bibitem{Pei:2005}
L. Pei, S. Hyun, J. F. Molinari, M. O. Robbins,
J. Mech. Phys. Solids {\bf 53}, 2385 (2005).

\bibitem{Yang:2008}
C. Yang and B. N. J. Persson, Phys. Rev. Lett. {\bf 100}, 024303 (2008).

\bibitem{Carbone09}
G. Carbone, M. Scaraggi and U. Tataglino,
Eur. Phys. J. E {\bf 30}, 65 (2009).

\bibitem{Campana10}
C. Campa{\~n}\'a, B. N. J. Persson, and M. H. M\"user,
J. Phys. Condens. Matter {\bf 23}, 085001 (2011).

\bibitem{Almqvist11}
A. Almqvist, C. Campa{\~n}\'a, N. Prodanov and B.N.J. Persson, J. Mech. Phys. Solids {\bf 59} 2355 (2011).

\bibitem{Robbins11}
S. Akarapu and T. Sharp and M. O. Robbins, Phys. Rev. Lett. {\bf 106}, 204301 (2011).

\bibitem{Carbone11}
G. Carbone and F. Bottiglione, Meccanica {\bf 46}, 557 (2011).

\bibitem{Pohrt12}
R. Pohrt and V. L. Popov, Phys. Rev. Lett. {\bf 108}, 104301 (2012).

\bibitem{Pohrt12PRE}
R. Pohrt, V. L. Popov and A. E. Filippov, Phys. Rev. E {\bf 86}, 026710 (2012).

\bibitem{Campana07}
C. Campa{\~n}\'a and M. H. M\"user, EPL {\bf 77}, 38005 (2007).

\bibitem{Campana08}
C. Campa{\~n}\'a, M. H. M\"user and M. O. Robbins, J. Phys.: Condens.
Matter {\bf 20}, 354013 (2008);
B. N. J. Persson, J. Phys.: Condens. Matter {\bf 20}, 315007 (2008).

\bibitem{Lorenz}
B. Lorenz and B. N. J. Persson, J. Phys.: Condens. Matter {\bf 21}, 015003 (2009)

%

\bibitem{Campana:2006}
C. Campa\~na and M.H. M\"user,
Phys. Rev. B {\bf 74}, 075420 (2006).

\bibitem{Pastewka:2012}
L. Pastewka, T.A. Sharp and M.O. Robbins,
Phys. Rev. B {\bf 86}, 075459 (2012).

\bibitem{Pharr1992}
G.~M. Pharr, W.~C. Oliver and F.~R. Brotzen, J. Mater. Res. {\bf 7}, 613 (1992).

\bibitem{Campana08b}
C. Campa{\~n}\'a, Phys. Rev. E {\bf 78}, 026110 (2008)

\bibitem{Barber13}
J.~R. Barber, Phys. Rev. E {\bf 87}, 013203 (2013)

\bibitem{Pelletier}
C.~G.~N. Pelletier, E.~C.~A. Dekkers, L.~E. Govaert, J.~M.~J. den Toonder and H.~E.~H. Meijer, Polym. Test. {\bf 26}, 949 (2007)

\bibitem{Dapp}
W.~B. Dapp, A. L\"ucke, B.~N.~J. Persson, M.~H. M\"user, Phys. Rev. Lett. {\bf 108}, 244301 (2012)

\bibitem{Buzio}
R. Buzio, C. Boragno, F. Biscarini, F.~B. de~Mongeot and U.~Valbusa, Nature Mater. {\bf 2}, 233 (2003)

\bibitem{luan05}
B.~Luan and M.~O. Robbins, Nature {\bf 435}, 929 (2005)

\bibitem{AmbaRao}
C.~L. Amba-Rao, J. Frankl. Inst. {\bf 287}, 241 (1969).

\bibitem{Johnson1985}
K.~L. Johnson, \emph{Contact Mechanics} (Cambridge University Press, 1985)

\bibitem{Hockney}
R.~W. Hockney, in: \emph{Methods in Computational Physics, Vol. 9}, pp.~135-211 (Academic Press, New York, 1970); see pp.~178-181

\bibitem{Nayak}
P.R. Nayak, J. Lubr. Technol. {\bf 93}, 398 (1971).

\end{thebibliography}
\end{document}